# The iEnvironment Platform:
## Developing an Open Science Software Platform for Integrated Environmental Monitoring and Modeling of Surface Water

Paulo Alencar, Donald Cowan, Doug Mulholland
David R. Cheriton School of Computer Science
University of Waterloo
Waterloo, Ontario, Canada
{palencar, dcowan, dmulholland}@uwaterloo.ca

*Abstract*—This paper describes the development of iEnvironment, an open science software platform that supports monitoring and modeling of aspects of surface water. The platform supports science and engineering research, especially in the context of the creation, sharing, analysis and maintenance of big and open data. In this era of big data, iEnvironment facilitates access to open data resources and research collaboration among science and research disciplines supported by computer scientists and software developers.

## I. INTRODUCTION

Surface water quality, quantity, availability, and management are crucial to society, and open and big data [1-4] about surface water is essential to environmental research. However, this data, although usually open, is distributed among all levels of government and NGOs, such as Conservation Authorities or Water Management Authorities and researchers. Further, the ability to model, monitor, and analyze complex interrelated surface-water related environmental processes is hampered by inadequate tools to cope with the data discovery, modelling, reuse, integration and sharing. In contrast, this research addresses how the existing disparate data can be accessed and shared easily and inexpensively while also capturing results from modelling runs for future use, so-called cumulative effects.

The scientific community recognizes this need for shareable scientific research platforms is an increasingly important issue, as much research conducted in specific, cutting-edge disciplines such as environmental science and engineering lacks shared community platforms [5-10].

This paper describes the development and extension of iEnvironment, an open science distributed data management platform and user gateway for integrated environmental monitoring and modelling (IEMM) related to surface water to support a number of research groups. The goal of the project is to: *improve significantly environmental science and engineering research on surface water by providing researchers with the ability to access and share environmental data and presentation facilities easily, while also developing best practices through sharing of best-of-breed monitoring, modelling and other analysis tools* [11-14]. The existing system is a sustainable collaborative research data platform for better access to surface water-related data and models, enabling researchers from multiple disciplines to collaborate and easily discover, access, combine and reuse data and models from multiple sources in order to perform novel forms of analyses. The extended platform will support new ways in which knowledge in the surface water sciences and engineering is created and applied to better understand water availability, quality, and dynamics.

The project involves new research teams consisting of surface water researchers in numerous disciplines, namely geomorphologists, hydraulic engineers, biologists, environmental scientists, as well as computer scientists and software developers. These experts acknowledge that the explosion of surface water-related digital information will transform the very nature of their research by supporting new research applications and greatly advance research in Canada. Indeed, in their various sub-fields, experiments, simulations, observations, instruments and surveys are generating exponentially growing amounts of all types of data. This information can be potentially combined to support new modes of discovery and lead to progress on a wide range of grand challenges, from climate change to natural storm management (e.g., floods, droughts), biodiversity impact (e.g., fisheries), water pollution prevention (e.g., from pollutants such as phosphorus), and water level management (e.g., river levels, receding shorelines). These water-related research challenges are clearly critical to Canada and to the world, especially in the light of recent water-related disasters that affected thousands of people and led to the loss of lives and billions of dollars in damage such as occurred in Walkerton water contamination (2000), Alberta and Muskoka Spring floods (2013) and algae in Lake Erie (2015, 2016).

iEnvironment is a technology platform that is lowering the barrier to environmental science and engineering researchers who need access to large amounts of environmental data in order to answer complex questions about the environment related to surface water. The platform is based on a framework-oriented architecture that is extensible, reusable and user-friendly. It provides an integrated set of services that individually and collectively can enable efficient, convenient and secure storage, publication, discovery, reuse and verification of data, by individual researchers, research teams, scientific collaborations, and the public at large. The platform also partially captures the processes that generate the data (e.g., metadata, models, tools), thereby supporting analysis, reuse and reproducibility. Further, shareable and discoverable data can enable collaboration and support repurposing for new approaches and interdisciplinary research. As a framework, iEnvironment can provide sustainable services defined within an extensible architecture which supports further tools and community-specific services to be added and enhance the framework over time. The research platform is based on open source software and will be freely accessible to the research community.

The current iEnvironment project is working with over 50 partners in universities, conservation authorities, other NGOs, governments at all levels and industry to advance environmental research and practice and develop the technology and processes behind environmental data sharing and IEMM. iEnvironment has been developed by the University of Waterloo Computer Systems Group (UWCSG) in consultation with many of these environmental partners to understand the needs of the environmental research community. iEnvironment is built using the Web Informatics Development Environment (WIDE) toolkit, an open toolkit developed by UWCSG.

## II. BACKGROUND

iEnvironment is supporting research teams leading projects that explore the next frontier of ground-breaking applications resulting from the integration of open or massive new types of data that can be used in integrated water monitoring and modelling for surface water. Examples of the proposed highly innovative applications that can enhance the existing research platform with tools and data include ones to support: (i) water quality and availability assessment; (ii) flowing water information systems including river networks; (ii) movement of physical, chemical and biological agents through the landscape; (iii) cumulative effects; (iv) modules for channel morphology, hydraulic, and biology; (v) components to map results, generate report cards, perform scenario analysis, and formalize cost/benefit and adaptive management strategies; (vi) community aquatic monitoring; (vii) biodiversity assessment (e.g., fish and other species, aquatic plants); environmental health assessment; (ix) monitoring and modelling stream behavior.

Specific research initiatives include projects involving: (i) the Community Aquatic Monitoring Program (CAMP), which investigates estuarine fish, invertebrates and water quality throughout the summer months; (ii) hydraulic engineering and fluvial morphology with a particular interest in river networks and dynamics using field measurements, physical experiments and computer simulations to understand river processes such as sediment transport and flow turbulence; (iii) conservation and restoration ecology, including studies involving population and community ecology studies on different habitats (e.g. prairies, forests, wetlands, shorelines, rivers) and taxa (e.g. plants, fungi, reptiles, arthropods, annelids); and resource limitation in streams and the role of organic matter, fish communities and ecosystem indicators for a changing system, stream benthos responses to riparian management, impact on amphibian communities, and the effects of forest practices.

*A. Making Use of Digital Infrastructures*

The proposed research data platform makes use of the digital infrastructures by connecting to the services supported by Compute Canada resources, which include store, compute and cloud capabilities. These resources are essential to the success of the project given the massive nature of some of the datasets and the analyses that need to be performed in surface water sciences and engineering research.

The research groups using this platform deal with a wide variety of problems to be solved that require different degrees of computational resources, namely storage, compute and cloud resources. First, the data to be stored and accessed by the platform includes historical and current weather data (e.g., from Environment Canada), water quality and nutrient source data, land use data, land shape data, remote sensing and field data. Present and future land use data is typically provided by municipalities and Watershed Management Authorities (WMAs), while drainage characteristics such as catchment and stream delineation are derived from digital elevation maps. Remote sensing and field data may come from WMAs, university researchers, professional consultants or government, while cumulative data is captured from previous studies. Second, the computational needs of the new research teams involve compute capabilities required by a number of models and tools, which involve, for example, point and non-point source nutrition, water balance, climate modelling, hydrology (e.g., snowmelt, stormwater management, flood control), biological stream health prediction, floodplain mapping, real-time flood forecasting, and cumulative effects. Predictive Models for Biological Stream Health through measurements of water course shape, fish and plant populations, and benthic data. The floodplain maps could be developed using digital elevation mapping acquired by Lidar-equipped aircraft or drones. The system could also use available in-stream/river bathymetry data. Real-time flood forecasting is another area where significant progress could be made, based on analyzing snowmelt and weather data, thereby anticipating the next Spring floods such as those that occurred in Muskoka and Calgary in Canada in 2013.

To address these storage and compute needs, the proposed platform makes use of Compute Canada resources, e.g., Graham (GP3) and Arbutus/west.cloud (GP1), using a programmatic data access API that abstracts away the details of specific ways used to store data and submit jobs to compute resources (e.g., clusters or clouds).

*B. Supporting New Research Applications*

The proposed research data platform will support and accelerate discovery in surface water sciences and engineering, an area in which Canada demonstrates world-class capabilities. Sustaining and growing research capabilities in this area are essential for Canada to tackle its own water problems and to meet global needs and opportunities.

The platform is supporting research teams consisting of surface water researchers in numerous disciplines, namely geomorphologists, hydraulic engineers, biologists, environmental scientists, as well as computer scientists and software developers. These experts acknowledge that the explosion of surface water-related digital information will transform the very nature of their research by supporting new research applications and greatly advance research in Canada. Indeed, in their various sub-fields, experiments, simulations, observations, instruments and surveys are generating exponentially growing amounts of all types of data. This information can be potentially combined to support new modes of discovery and lead to progress on a wide range of grand challenges, from climate change to natural storm management (e.g., floods, droughts), biodiversity impact (e.g., fisheries), water pollution prevention (e.g., from pollutants such as phosphorus), and water level management (e.g., river levels, receding shorelines). These water-related research challenges are clearly critical to Canada and to the world, especially in the light of recent water-related disasters that affected thousands of people and led to the loss of lives and billions of dollars in damage such as occurred in Walkerton water contamination (2000), Alberta and Muskoka Spring floods (2013) and algae in Lake Erie (2015, 2016).

The research teams are supported in research tasks related, among others, to storing and accessing data, protecting data through access control mechanisms, developing modelling and monitoring models and algorithms, processing algorithms, and recording provenance data and results. Specific projects the new research teams will be developing based on the research platform focus, for example, on:

- Improving disaster warning by developing scientific knowledge, monitoring and modelling techniques, and national forecasting capacity to predict the risk and severity of extreme events;
- Understanding and prediction of hydrological and terrestrial processes and watershed hydrology and how processes and systems evolve and interact under a changing climate;
- Developing decision support systems to improve planning decisions around urbanizing watersheds;
- Developing improved models, including predictive and diagnostic modelling system development and deployment for hydrology, water quality and water resources;
- Preserving ecosystem health and conservation, and developing better models to assess change in human/natural land and water systems;
- Developing river monitoring strategies to assess physical adjustment of river systems in response to changes in flow and sedimentological regimes;
- Developing big data approaches for surface water problems, using sensors, instrumented river basins, and data analysis systems;
- Investigating complex system modelling and analyses that involve the interactive dynamics in human-natural coupled systems;
- Developing novel methods to deal with cumulative effects, that is, effects on the environment which are caused by the combined result of past, current and future activities;
- Undertaking knowledge mobilization for decision support, including the facilitation of communities of practice, stakeholder engagement with science, visualization and decision theatres, development of place-based solutions for climate adaptation, and evidence-based decision making.

III. SYSTEM ARCHITECTURE

A. High-Level Architecture

iEnvironment is a web-based, distributed data management platform and user gateway for integrated environmental monitoring and modelling (IEMM) related to surface water. The platform supports access to and sharing of data from multiple distributed heterogeneous data sources, as well as monitoring and modelling models and tools to enable researchers to perform complex analysis easily and efficiently on the integrated data. By abstracting away details about the complexities and variations of the data sources and the compute and storage jobs, the platform supports hiding these technical details and enabling researchers to perform complex and computationally expensive data analyses. The platform also captures results from previous analyses so that future analyses can build upon previous results showing cumulative effects.

iEnvironment shown in Figure 1, is a multi-tiered platform composed of three main layers, namely: (i) the user interface layer; (ii) the application layer; and (iii) the data layer. The following diagram outlines the hardware and software components of the proposed solution, and identifies using numbers (1-8) the parts of the system that have to be improved, modified or added to support multiple research teams better.

The platform is accessible through a standard web browser that provides external access to data users (which can be human users or other applications/platforms). Through the Internet, data users access the user interface layer by connecting to an administration server that supports security (i.e., authentication and authorization). The main components of the user interface (UI) layer are web and mapping servers, user interface tools and a user access control service. The application layer involves an application server, monitoring and modelling application components and a process management component, which can capture process-related information for reproducibility purposes. This layer can access remote computational resources (i.e., store, compute and cloud resources) provided by Compute Canada. The data layer involves data management capabilities, including data management tools, data and metadata storage, and can have access to external servers and external cloud servers, e.g., Compute Canada or Amazon Web Services (AWS). The Internet connection with the data providers supports data access control and security/privacy services. The boxes numbered 1-2 and 4-7 indicate the components that will be enhanced or modified in the system. The boxes numbered 3 and 8 indicate the added components. Beyond the existing application components supported by iEnvironment, novel applications and related data will be added to support new research groups (box 3).

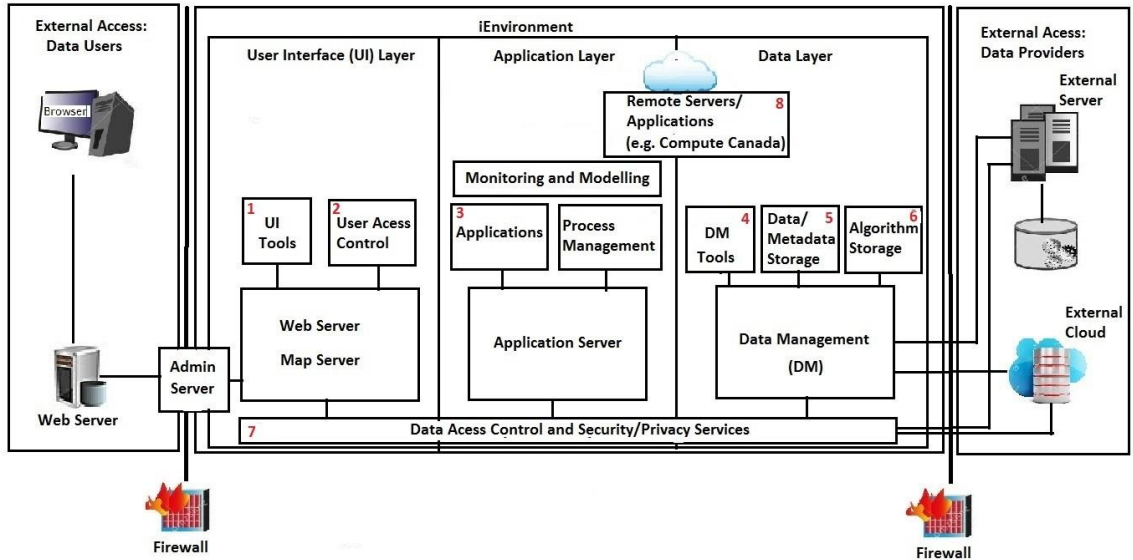

**Figure 1. iEnvironment: High-Level Architecture.**

The current web and cloud-based data and application management platform currently contains two large applications, one supporting monitoring/modelling and the other one only modelling. There are also a number of monitoring, modelling tools, a comprehensive user interface and access control tools. The platform also supports tools to capture and maintain data bases from multiple sources. The system also supports mobile devices for use with data entry and display. The monitoring/modelling application is called the Flowing Waters Information System (FWIS) and the modelling application is the 5th version of the CANadian Watershed Evaluation Tool (CANWET™/5). The platform also contains a modified version of the R tool that supports statistical analysis of server-based databases. The user interface contains a set of tools to produce reports, graphs and charts and maps. These tools can be used with each system or to overlay results from multiple systems. The user interface also contains the first version of a tool that supports automated generation of scripts for the acquisition and maintenance of data acquired for the internal databases. There is also an underlying project management system that supports the recording of projects including: geographic location and type of data collected. This project management tool can be easily augmented to support the addition of other data.

FWIS was developed in partnership with the Ontario Ministry of Natural Resources and Forestry and several conservation authorities. The purpose of FWIS is to provide researchers, municipalities and resource agencies with access to collective stream fisheries data in the lower Great Lakes basin to protect sensitive fish habitat. FWIS provides reasonably up-to-date and comprehensive fish species habitat and stream flow information. As well FWIS supports 17 study types including: Benthics, Fish, Channel Morphology, Channel Stability and Discharge. FWIS offers an approach for identifying where and what data has been collected, who collected the data, and which protocols were used. This information will facilitate better science development, state of resource reporting, monitoring and data sharing. CANWET™/5 is a web and cloud-based implementation of an earlier desktop version that has been used to model watersheds for several Ontario Conservation Authorities (CA) including Lake Simcoe, Nottawasaga and Mississippi Valley. CANWET™/5 is currently being modified to support multiple what-if scenarios and is in use by two organizations listed in the section "CANWET™ Users." The CANWET™ platform is a modelling software suite designed to inform decision making around river basin and watershed management; water supply and wastewater treatment infrastructure; food security; and, climate change adaptation. CANWET™ can quickly estimate daily water balance, nutrient, erosion sediment, and bacteria loadings from map-based input data adjusted for potential best management practices (BMPs), thus making CANWET™ a powerful decision support system. CANWET™ has been used for Assimilative Capacity Studies; Watershed and Sub-watershed Studies; Master Drainage Plans; Infrastructure Planning; and Source Water Protection Studies.

iEnvironment has been developed in a framework-oriented architecture style, which makes it easier for components to be extended or reused by relying on design patterns and explicitly providing software extension points. The platform is stable and the successful development of previous applications such as CANWET™/5 and FWIS have heavily relied on its reuse and extendibility capabilities.

*B. Architectural Components*

In Figure 2 we introduce a high-level architecture diagram of the major functional components of the proposed software, describe how each component interacts with each other, and identify using numbers (1-8) the components that will be improved, modified or added to the existing system.

The following services are being improved, modified or added to the system (1-8). The boxes numbered 1-2 and 4-7 denote the architectural components that will be improved or modified to support the new research teams:

- (1) Improved user interface services and their APIs, such as dashboards and administration tools to support project management (e.g., for individuals, groups, organizations using the data);
- (2,7) An expanded login procedure as well as user and data access control services and their APIs to support access to the additional data and tools;
- (4) Improved data management service tools and their APIs to support data search, publication, access, and other relevant services;
- (4) Improved data management services to support data ingestion (e.g., data import/export), that is, the automated transfer and maintenance of data from multiple external sources into an internal database, data adaptation, migration and other relevant services;
- (5,6) Data, metadata and algorithm storage services and their APIs.

The other numbered boxes (3, 8) denote the new application components:

- (3) New service application components and their APIs to enable easy access to surface water monitoring and modelling applications and their relevant datasets, which include tools to model water quality, quantity and dynamics (e.g., rivers, biodiversity);
- (3) An improved service and its API that supports the ability to create working sets so that researchers can experiment with what-if scenarios without impacting existing data and processes;
- (3) An improved notification (publish-subscribe) service, which can use any attribute in the data or metadata models (e.g., for notifying about changes in data, metadata, algorithms, projects, teams);
- (8) Instantiated access services to the Compute Canada resources (e.g., store, compute, cloud), especially the GP3 cluster, one of the three largest supercomputing cluster that is being installed at the University of Waterloo, Waterloo, Ontario, Canada.

The platform is currently built on a Windows platform and allows connections to other systems (e.g., Unix-like systems) using ODBC, JDBC, native database connections, Samba and other mechanisms, thus supporting data and file sharing between these two platforms. The platform has also been run directly on a Linux/Apache host, although some modelling facilities are not currently available for Linux. The client rich user interface that provides external data access to iEnvironment users is built through a web browser for user access in a modular, flexible, framework-oriented style, using PHP, JavaScript, HTML5, CSS, JSON, and REST-Ajax. It provides support for web, map and mobile access through system-specific visual dashboards, administration tools, and customized user access control. The user interface supports mapping tools through MapServer, a mapping server based on Web Map Service (WMS), which is a standard protocol for serving map images (over the Internet) that are generated using data from a geospatial database or SHP file. The WIDE toolkit includes a WMS-compatible map viewer or, alternatively, other WMS-compatible tools can be used (e.g., Leaflet, Google Earth). Other platforms and applications can access the iEnvironment platform using RESTful services and cURL APIs. cURL which is a "Client for URLs", supports transmitting and receiving data using all standard protocols, including HTTP/HTTPS. It can also be used to test web applications from the automated test framework of the WIDE toolkit. cURL can be accessed from most popular programming environments, including R and RStudio to create a high-quality GUI system, supporting the creation of rich graphical applications. The application layer is based on PHP, an object-oriented scripting language that is very flexible and can run on Windows, Linux (and other Unix-like systems), and macOS/OS X, and support connecting to application and data servers as well as to databases such as SQL/Anywhere, MS SQL Server, Oracle or MySQL. PHP also integrates well with systems that store and process big data sets and supports scalability by making it easier to add more servers or computing nodes as the project grows. PHP can be used to access the databases, generate or transform XML, which can be transmitted to and rendered in the user interface browser if required. Existing systems that have been built as platform extensions include CANWET™/5, FWIS, and R-based tools for data analyses. The data layer is supported by data management tools and data/metadata/algorithm storage databases (i.e., whichever database is selected for a host) and retrieval mechanisms. Functional components including components supporting searching, publishing and accessing datasets are provided. Data acquisition components also includes a data ingestion (e.g., import/export) component, a data adaptation/migration component, a data access control/security/privacy component and other services.

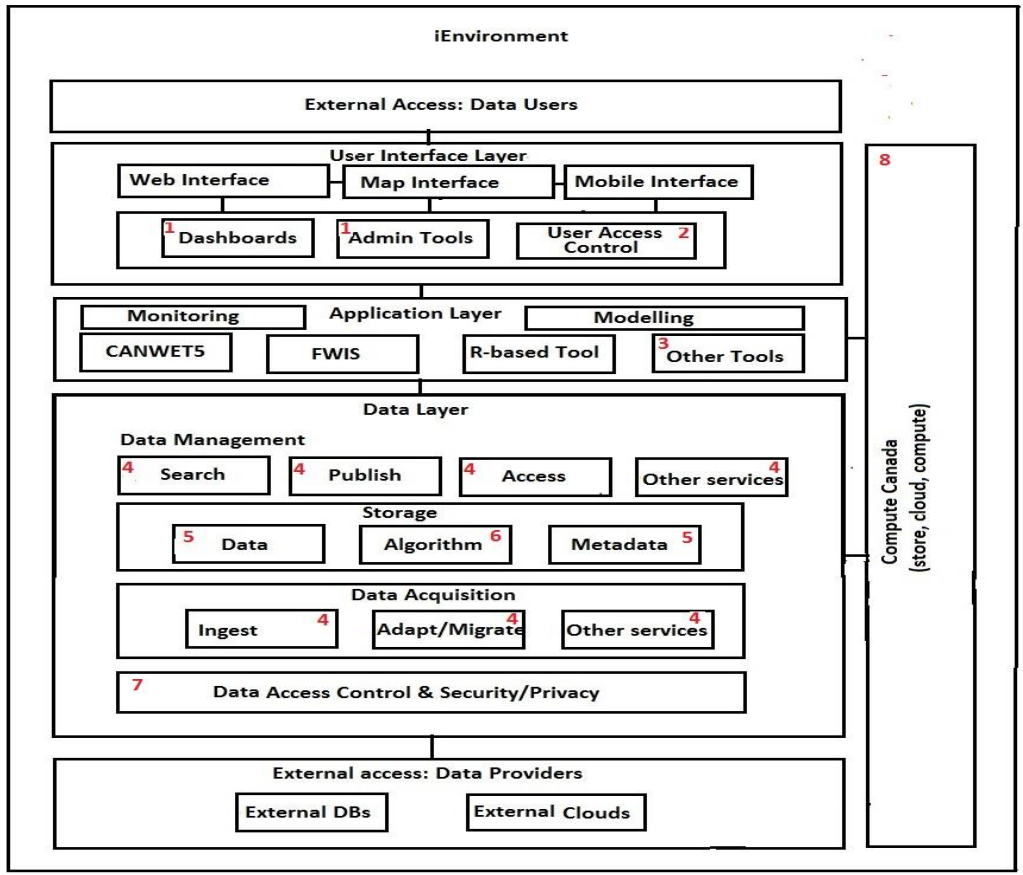

**Figure 2. Architectural Components.**

Access to external data providers (e.g., external servers, Compute Canada resources) can be done through PHP or via SQL database-to-database connections, which support connecting with data sets in remote databases, and accessing files and other media objects such as images.

*C. Coping with Big and Open Data*

Four serious problems facing the environmental community besides locating sources of data are:

i meeting the challenge of using big data in environmental research;
ii coping with environmental big data;
iii capturing open data in a useable form;
iv capturing historical data.

Experts acknowledge that access to big data related to surface water will transform the very nature of their research by supporting new research applications. Indeed, in their various sub-fields, experiments, simulations, observations, surveys, new instrumentation have the potential to generate huge amounts of all types of data. This information can be combined to support possible new modes of discovery and lead to progress on a wide range of grand challenges, from climate change to natural storm management (e.g., floods, droughts), biodiversity impact (e.g., fisheries), water pollu- tion prevention (e.g., from pollutants such as phosphorus), and water level management (e.g., river levels, receding shorelines). One big challenge will to be see how far the iEnvironment framework can adapt and evolve to cope with this deluge of big data.

Environmental data is big data but most monitoring or modelling considers a watershed or sub-watershed area where the amount of data is large but manageable. In our experience of 25 years working on environmental information systems we have found that relational database technology works well for this form of geographic big data.

A large amount of environmental data is becoming open. However, the data is rarely useable in its available format and must be imported into a database system for reasons of organization and performance. Of course the data is not imported once, but must be kept up to date.

Finally, there is historical data, which may have been collected for decades. Recently, in the last 20 to 30 years, the data may have been kept in machine readable format such as spreadsheets or simple database structures.

Before that there were written records. Even if the data was machine readable, the data was often stored on a standalone computer or in a desk drawer on some media.

All these factors must be considered as iEnvironment is being produced. The platform must support data ingestion tools to capture and maintain data bases from machine- readable sources and mobile devices for data entry from the field. Such tools have been developed and are being improved based on experience. Of course, data recorded on paper must eventually be entered manually or by scanning. Often an organization will enter the data into its own database and the iEnvironment platform must then capture the data from there using the data ingestion tools.

*D. Ongoing Modification Summary*

The modifications that are being made to the existing software platform involve the following components, which will be enhanced or added to the existing platform to support the new research teams:

- Improved login/reporting procedure

This procedure will provide support for the new research team reporting requirements as well as the basic login features.

- Instantiated project support tool API

This service will provide automated support for project management, supporting mechanisms to manage teams, projects, team members, and productivity metrics.

- Instantiated user and data access control API

This improved user and data access control component will allow customized access control based on multiple teams, projects, team members, and their various data sources, algorithms, and modelling and monitoring tools.

- Improved data storage service API

This component will allow database support for the data sources to be added by the new research teams as well as various retrieving mechanism for the specific data.

- Improved metadata storage service API

This component will support metadata databases to support provenance data (e.g., data creation, changes/updates, computational jobs, job duration) as well as retrieving mechanisms.

- Improved algorithm storage service API

This component will support storing a wide variety of analysis algorithms in databases and provide retrieving mechanisms.

- New monitoring and modelling application component APIs

These new components are the discipline-specific components that have been or are being developed by the new research teams, and will be made available in the platform to support their data and analysis requirements needs.

- Instantiated search service API

This component will provide automated support for searching datasets, analysis algorithms, metadata, and other information relevant to the new research teams.

- Instantiated publication service API

This component will support the publication of new datasets, algorithms and tools by the new research groups.

- Instantiated data, algorithm and tool access API

This component provides access to data, algorithms and tools provided in external sources (external servers, databases).

- Other customized data management service APIs

These components constitute support for other data management services provided by the research platform (e.g., data consistency, data integrity components).

- Improved data ingestion API

This component supports (semi-)automated data ingestion procedures, which can be used to facilitate functions such as data import and data export from external data sources.

- New access to Compute Canada resources API

This new component will support accessing Compute Canada resources, especially GP3, one of the largest supercomputing clusters in Canada.

- New what-if scenario analysis API

This component will support experimentation through what-if scenarios using the new monitoring and modelling application components developed by the new research teams.

- New notification (publish-subscribe) service

This service provides notifications based on the data, algorithm and metadata models, and can support notifications that include notifications related to data changes/updates, provenance data changes, model changes, as well as project and team changes.

- Instantiated dashboard service API

This component supports customized data visualization, supporting various formats (e.g., tables, charts, graphs, maps), and can provide customized user interaction.

As can be inferred from this list of deliverables, the software development will involve enhancing existing platform components as well as adding new ones to the platform. First, data and application components that have been or are being developed by the new research groups will be added to the platform. In addition, a new component needs to be added to support access to Canada Compute resources, and a notification component will be added to provide customized notifications to the user groups. Finally, a new component needs to be added to support what-if scenario analysis related to the added monitoring and modelling applications components provided by the new research groups.

The project deliverables developed in this project will be open source and freely accessible to the research community. They are operated by the Computer Systems Group at the University of Waterloo (UWCSG), and are available under the "3 Clause BSD" licence.

*E. The Platform's Compute and Storage Strategy*

The research teams are supported through a compute and storage allocation strategy that will be detailed at the starting phase of the project and has the following features: (i) new research teams have their data, algorithms and tools, managed by the research platform project management tool, and can define appropriate access rights to individual team members or sub-groups according to their specific team structures; (ii) new research teams can use personal allocations of resources on Compute Canada for supporting their applications or use allocations allocated to the research platform itself in terms of storage, compute and cloud services; (iii) the new research teams can use their allocated resources for production tasks, while the software development team will use the resources for development, testing and other pre-production tasks; (iv) the research platform can be designed to use resources opportunistically and should ideally be able to exploit idle cycles on Compute Canada and potentially increase the scale of computing resources; (v) compute and storage resources will be monitored in the various cases, including servers/workstations/clouds; and (vi) the Lead Software Developer will be directly involved in compute and storage resource allocation decisions.

The new research teams will take advantage of Compute Canada resources, especially Graham (GP3) and Arbutus/west.cloud (GP1). GP3 will be the main focus of the allocation strategy and is a heterogeneous cluster, suitable for a variety of workloads.

## IV. FEATURE CUSTOMIZATION AND EXTENSIONS

The proposed research platform software design allows for future customization and extension of functionality to meet the needs of additional research teams and new research applications by providing a modular, flexible, framework-oriented architecture style that explicitly defines reuse components and extension points. A framework is defined as a set of pre-built software building blocks that developers can use, extend, or customize to construct specific applications. The development team has been working with architectures that aim at reuse and extensibility for decades. In fact, this is one of the areas of expertise of the Principal Investigator.

Regarding the addition of research teams, the software design was defined so that the research platform would support collaboration in different ways: (i) the existing system allows the addition of new research teams by including a project reporting mechanism that supports the addition of new organizations, teams, and individuals in various roles; (ii) the existing system supports access control mechanisms that allow the definition of structure of new research groups, including data protection and privacy directives, and also supports communication and coordination efforts by allowing persons with different expertise and access rights to cooperate; (iii) new research groups can create specialized views for their data and tools, and annotate and change data and tools in their contexts without interfering with other users; (iv) the system supports cooperative functions such as common data annotation, so that the data can be augmented and shared within a group and new content or tools can be created cooperatively; (vi) research team data is reused within the communication tools so that individuals in a specific group have access rights compatible with the data managed by the group; (v) the ability to store and link data sources and algorithms within groups also provides a way to support new knowledge creation as well as new tool processing; and (vi) the system design supports leveraging the domain experts' experience within the new research teams.

Further, regarding the addition of new research applications, the software design is defined to support reuse and extendibility in various ways: (i) supporting a framework-oriented, pattern-based architecture that is highly reusable and extendible; (ii) allowing convenient (easy-to-perform) data storage and access by adopting user-friendly and intuitive forms-based interfaces; (iii) supporting reuse and extension of data, algorithms and tools by relying on framework-style extension points; (iv) allowing flexible integration of data and applications by supporting standardized input and output interfaces; (v) providing extension points that support different types of model and code customization, e. g., new data and application properties, new types of diagrams for visualization, new menu entries, extension (or variation) code, and extension package modules; (vi) system extensions can be provided by new research teams or other end users, which can safely contribute any functionality in a controlled way (e.g., by restricting access control rights); and (vii) the system framework design style often leads to benefits such as reduced maintenance costs, and higher productivity.

## IV. CONTRIBUTIONS OF iENVIRONMENT

The iEnvironment platform has been providing a wide variety of contributions. First, it enhances opportunities for collaborative knowledge creation and innovation within Canada's research and education communities through the maintenance and development of the CANARIE Network and related tools and services by:

- contributing to the development of new common and rationalized data sources, services and tools that are essential for the collaborative research initiative within surface water sciences and engineering;
- allowing surface water researchers in numerous disciplines, namely geomorphologists, hydraulic engineers, biologists, environmental scientists, as well as computer scientists and software developers, to have a platform that will accelerate their research discovery by supporting reuse and integration;
- providing a platform that will support new data and models, thereby allowing new questions to be asked that will lead to progress on a wide range of grand challenges, from climate change to natural storm management (e.g., floods, droughts), biodiversity

- impact, water pollution prevention, and water accessibility management;
- making the research platform available to other research teams, government agencies, industries and not-for-profit organizations;
- providing a strong student and workforce training and education component.

Second, the platform leads to an expansion of the research and education community's access to and utilization of the CANARIE Network and the availability of tools and programming that increase the effectiveness of its use by:
- enabling enhanced retrieval, ingestion, archiving and searching of structured and unstructured content from distributed, disparate sources.

Finally, the platform enables the creation of innovative Information and Communications Technology (ICT) products and services and accelerates their commercialization in Canada by:
- presenting an economic opportunity in that the digital data revolution is global, and thus the market for new data-driven water products and services is immense;
- supporting research teams that are well positioned to work with industry and government partners throughout Ontario's growing water and ICT industrial organizations, federal and provincial governments, and the consulting sector to take full advantage of the economic opportunities that may arise as a result of the research platform capabilities;
- supporting researchers who not only are leaders in their fields, but also have ongoing collaborative efforts with numerous water-related industries and government agencies and strong records of technology transfer.

*A. Advancing Research Software Development*

The proposed platform is greatly advancing existing capabilities in research software development. The development group at the University of Waterloo has an excellent track record of developing and deploying software for provincial, national and international projects. The group has already developed several systems by reusing and extending the existing research platform, including CANWET™/5, FWIS, project management support for collaborative efforts, and R-based tools for data analyses.

From the technical standpoint, the availability of the proposed software platform will: (i) augment the software development capability related to one of the main strategic domains in Canada, namely water-related sciences; (ii) support the development of data sources and discipline-specific software in areas in which Canada produces world-class research; (iii) enhance the new research groups' capabilities in the creation of data and tools and will generate additional opportunities across both the water scientific communities and water industries; (iv) advance research software development capabilities with a leading-edge research platform that will support the automation of research-related tasks (e.g., data access and storage, data analysis, modelling, monitoring) that will impact the way water sciences and engineering applications are developed; (v) develop software modelling and monitoring technologies that could be reused or adapted to other scientific areas or industries (e.g., healthcare, precision agriculture, energy), allowing users to increase productivity, reduce costs and become more competitive and provide tools and processes to access open data more easily from multiple sources.

From the standpoint of the new research teams, the project will: (i) enhance inter-group collaboration and coordination of efforts to avoid work duplication and minimize the "not invented here" phenomenon; (ii) build on the world-class expertise and capabilities that have been developed over the past several years in producing IEMM software and produce benefits to Canada; (iii) expand the expertise in surface water sciences and engineering; and (iv) support the development of testbeds for new software technologies.

From the viewpoint of human resources, the project will: (i) help to develop additional highly qualified personnel in surface water sciences and engineering that are knowledgeable in data-centric software technology to support research; (ii) encourage the creation of new software development projects and jobs related to the research data platform efforts; (iii) provide growth opportunities for the collaborating scientific, government agencies, industrial companies, and not-for-profit organizations, and enable them to increase their level of software development; (iv) enable highly qualified personnel to develop software for applications in areas involving open and big data, which can provide job opportunities that are in high demand; (v) provide world-class training opportunities for students and faculty; and (vi) help to attract more students to the areas supported by the research platform.

From a national strategic standpoint, the proposed project will: (i) provide research development capabilities for established national, or indeed, international networks of researchers; (ii) help to improve the capabilities, competitiveness and performance of the Canadian research community in the addressed areas; (iii) improve Canada's active base of surface water research in numerous universities, research institutes, government agencies, industries and other organizations; and (iv) help in producing one of the most important outcomes of CANARIE's contributions to Canada's national research development capacities.

V. CONCLUSIONS

This paper describes the development of iEnvironment, an open science platform (and its components) that enables new research teams to access open and big data sources, and produce tools and applications. This open science software platform deliverables will significantly advance research for many reasons as they will: (i) augment the national research capability in the area of expertise of the new research teams, an area in which Canada is recognized as an international leader; (ii) help to access data, and produce tools and applications that will greatly benefit Canada's water researchers in that they will be enabled to ask complex questions based on multiple and heterogeneous data sources that were previously impossible to be addressed; (iii) lead to novel research opportunities based on the synergy among novel data sources, monitoring and modelling tools, the new research applications and the research teams; (iv) facilitate the development of new research-based tools and applications based on the reusable and extendible aspects of the research platform; (v) allow the research teams to increase their productivity, reduce their costs and become more competitive; (vi) support collaborative efforts within and among research groups interested in related topics; (vii) allow Canadian researchers to focus on developing their models and applications instead of on the development of tools for data storage, sharing and processing; (viii) provide automated support for the development of new modes of discovery, as well as novel computational models and analyses; (ix) enable an established national, or indeed, international, network of researchers; and (x) improve the research capabilities and competitiveness of the Canadian surface water research community.

ACKNOWLEDGMENTS

The authors thank CANARIE, the Ontario Ministry of Natural Resources and Forestry, the Ontario Research Excellence Fund (ORF), the Natural Sciences and Engineering Research Council (NSERC), the Centre for Community Mapping, Greenland International Consulting, FedDev, SAP Canada, and the United Nations University Institute for Water, Environment and Health (UNU-INWEH) for support.